\documentclass{article}
\usepackage{spconf,amsmath,graphicx}
\usepackage[detect-all]{siunitx}
\usepackage{mathtools}
\usepackage{comment}
\usepackage{multirow}
\usepackage{bbm}
\usepackage[subrefformat=parens]{subfig}
\usepackage{threeparttable}
\usepackage{booktabs}
\usepackage{amssymb}
\usepackage{hyperref}

\usepackage{xspace}
\makeatletter
\DeclareRobustCommand\onedot{\futurelet\@let@token\@onedot}
\def\@onedot{\ifx\@let@token.\else.\null\fi\xspace}
\def\eg{\emph{e.g}\onedot} 
\def\ie{\emph{i.e}\onedot}

\makeatother

\def\equationautorefname~#1\null{(#1\null)}

\newcommand{\vect}[1]{\mbox{\boldmath $#1$}}

\newcommand{\trans}[1]{#1^\mathsf{T}}

\def\appendixautorefname~#1\null{~#1 \null}

\makeatletter
\newcommand{\figcaption}[1]{\def\@captype{figure}\caption{#1}}
\newcommand{\tblcaption}[1]{\def\@captype{table}\caption{#1}}
\makeatother

\title{Multi-Channel End-to-End Neural Diarization\\with Distributed Microphones}
\name{%
Shota Horiguchi$^{\star}$ \quad
Yuki Takashima$^{\star}$ \quad
Paola Garc\'{i}a$^{\dagger}$ \quad
Shinji Watanabe$^{\ddagger}$ \quad
Yohei Kawaguchi$^{\star}$
}
\address{%
$^{\star}$ \text{Hitachi, Ltd., Japan} \\
$^{\dagger}$ \text{CLSP \& HLTCOE, Johns Hopkins University, USA}\\
$^{\ddagger}$ \text{Carnegie Mellon University, USA}
}
\begin{document}
\ninept

\abovedisplayskip=3pt
\belowdisplayskip=3pt

\setlength\floatsep{10pt}
\setlength\textfloatsep{10pt}
\setlength\intextsep{10pt}
\setlength\abovecaptionskip{3pt}
\setlength\belowcaptionskip{3pt}
\setlength\abovetopsep{5pt}
\setlength\dbltextfloatsep{9pt}

\aboverulesep=0.25ex 
\belowrulesep=0.5ex 
\maketitle
\begin{abstract} 
Recent progress on end-to-end neural diarization (EEND) has enabled overlap-aware speaker diarization with a single neural network.
This paper proposes to enhance EEND by using multi-channel signals from distributed microphones.
We replace Transformer encoders in EEND with two types of encoders that process a multi-channel input: spatio-temporal and co-attention encoders.
Both are independent of the number and geometry of microphones and suitable for distributed microphone settings.
We also propose a model adaptation method using only single-channel recordings.
With simulated and real-recorded datasets, we demonstrated that the proposed method outperformed conventional EEND when a multi-channel input was given while maintaining comparable performance with a single-channel input.
We also showed that the proposed method performed well even when spatial information is inoperative given multi-channel inputs, such as in hybrid meetings in which the utterances of multiple remote participants are played back from the same loudspeaker.
\end{abstract}
\section{Introduction}
\label{sec:intro}
Meeting transcription is one of the largest application areas of speech-related technologies.
One important component of meeting transcription is speaker diarization \cite{anguera2012speaker,park2021review}, which gives speaker attributes to each transcribed utterance.
In recent years, many end-to-end diarization methods have been proposed \cite{fujita2019end1,horiguchi2020endtoend,medennikov2020targetspeaker,horiguchi2021encoder} have achieved comparative accuracy to that of modular-based methods \cite{landini2022bayesian,park2020auto}.
However, many attempts have been made based on single-channel recordings, where no spatial information is available.
Some meeting transcription systems are based on distributed microphones \cite{araki2017meeting,araki2018meeting,yoshioka2019meeting,horiguchi2020utterance}, which enables the flexibility of recording devices and a wide range of sound collection.
If we can improve diarization accuracy by extending the diarization methods to distributed microphone settings, it will be compatible with those systems.

Even if multi-channel inputs are given, diarization methods that heavily rely on spatial information are sometimes inoperative.
The best examples are direction-of-arrival (DOA) based diarization \cite{araki2008doa,ishiguro2011probabilistic}.
Due to COVID-19, meetings are now often held remotely or in a hybrid version of in-person and virtual gatherings.
In hybrid meetings, remote attendees' utterances are played via one loudspeaker, and DOA is no longer a clue to distinguish these speakers.
To cope with this situation, spatial information needs to be properly incorporated into speaker-characteristic-based speaker diarization.

In this paper, we propose multi-channel end-to-end neural diarization (EEND) that is invariant to the number and order of channels for distributed microphone settings.
We replaced Transformer encoders in the conventional EEND \cite{horiguchi2020endtoend,horiguchi2021encoder} with two types of multi-channel encoders.
One is a spatio-temporal encoder \cite{wang2020neural,wang2021continuous}, in which cross-channel and cross-frame self-attentions are stacked.
It was reported in the context of speech separation that the encoder performs well when the number of microphones is large but degrads significantly when the number of microphones is small \cite{wang2020neural}.
The other encoder is a co-attention encoder, in which both single- and multi-channel inputs are used and cross-frame co-attention weights are calculated from the multi-channel input.
There are only cross-frame attentions; thus, its performance does not heavily depend on the number of channels.
We further propose to adapt multi-channel EEND only with single-channel real recordings without losing the ability to benefit from spatial information given a multi-channel input during inference.
We show that the proposed method can utilize spatial information and outperform the conventional EEND.

\section{Related work}
Some multi-channel diarization methods are fully based on DOA estimation \cite{araki2008doa,ishiguro2011probabilistic}, but assume that different speakers are not in the same direction, thus are not appropriate for hybrid meetings.
Therefore, spatial information needs to be incorporated with single-channel-based methods as in \eg \cite{anguera2007acoustic}.
Another possible approach is to combine channel-wise diarization results by using an ensemble method \cite{stolcke2019dover,raj2021doverlap}, but it does not fully utilize spatial information.
Some recent neural-network-based diarization methods utilize spatial information by aggregating multi-channel features.
For example, online RSAN \cite{kinoshita2020tackling} uses inter-microphone phase difference features in addition to a single-channel magnitude spectrogram. However, the number of channels is fixed due to the network architecture, making the method less flexible. Moreover, phase-based features are not suited for distributed microphone settings, in which clock drift between channels exists.
Multi-channel target-speaker voice activity detection (TS-VAD) \cite{medennikov2020targetspeaker} combines embeddings extracted from the second from the last layer of single-channel TS-VAD.
Although it is flexible in terms of the number of channels because an attention-based combination is used, it requires an external diarization system that gives an initial i-vector estimation for each speaker.

If we broaden our view to speech processing other than diarization, there are several methods for neural-network-based end-to-end multi-channel speech processing that are invariant to the number of channels, \eg speech recognition \cite{ochiai2017multichannel,wang2019stream,chang2021multichannel}, separation \cite{luo2019fasnet,luo2020endtoend,wang2020neural,furnon2021distributed,wang2021continuous}, and dereverberation \cite{wang2020multi}.
Many  use attention mechanisms to work with an arbitrary number of channels.
Our proposed method also uses attention-based multi-channel processing.

\section{Conventional single-channel EEND}
\subsection{Formulation of EEND}
In the EEND framework, $S$ speakers' speech activities are jointly estimated.
Given $F$-dimensional acoustic features for each $T$ frames $X\in\mathbb{R}^{F\times T}$, we first apply a linear projection parameterized by $W_0\in\mathbb{R}^{D\times F}$ and $\vect{b}_0\in\mathbb{R}^D$ followed by layer normalization \cite{ba2016layer} $\mathsf{LN}$ to obtain $D$-dimensional frame-wise embeddings
\begin{align}
    E^{(0)}=\mathsf{LN}\left(W_0X+\vect{b}_0\trans{\vect{1}}\right)\in\mathbb{R}^{D\times T},\label{eq:change_dimension}
\end{align}
where $\vect{1}$ is the $T$-dimensional all-one vector.
It is further converted by $N$-stacked encoders, where the $n$-th encoder converts frame-wise embeddings $E^{(n-1)}$ into the same dimensional embeddings $E^{(n)}$:
\begin{align}
    E^{(n)}=\mathsf{Encoder}\left(E^{(n-1)}\right)\in\mathbb{R}^{D\times T}.\label{eq:encoder}
\end{align}
Finally, the frame-wise posteriors of speech activities for $S$ speakers are estimated.
In this paper, we used EEND-EDA \cite{horiguchi2020endtoend,horiguchi2021encoder}, with which the speaker-wise attractor $B$ is first calculated using an encoder-decoder based attractor calculation module (EDA) and then the posteriors $Y$ are estimated as
\begin{align}
    B&=\mathsf{EDA}\left(E^{(N)}\right)\in\mathbb{R}^{D\times S},\label{eq:attractors}\\
    Y&=\sigma\left(\trans{B}E^{(N)}\right)\in\left(0,1\right)^{S\times T},\label{eq:posteriors}
\end{align}
where $\sigma\left(\cdot\right)$ is the element-wise sigmoid function.
A permutation-free objective is used for optimization as in previous studies \cite{fujita2019end1,horiguchi2020endtoend,horiguchi2021encoder}.

\subsection{Transformer encoder}
\begin{figure}[t]
    \centering
    \subfloat[][Transformer encoder]{\includegraphics[width=0.9\linewidth]{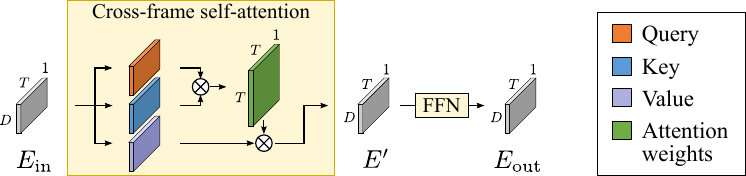}\label{fig:transformer}}\\
    \subfloat[][Spatio-temporal encoder]{\includegraphics[width=0.9\linewidth]{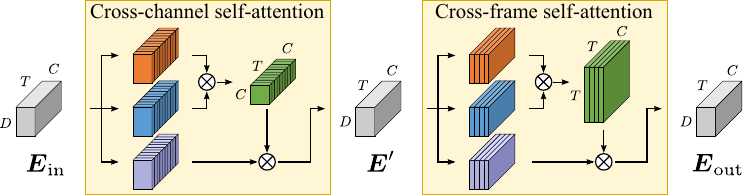}\label{fig:spatiotemporal}}\\
    \subfloat[][Co-attention encoder]{\includegraphics[width=0.9\linewidth]{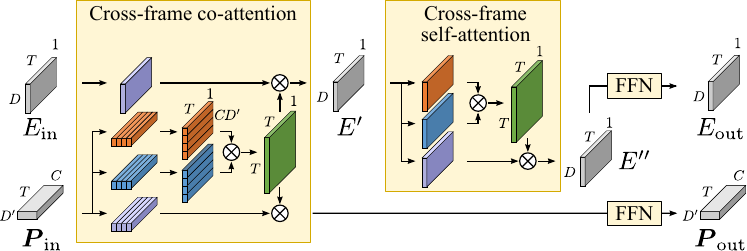}\label{fig:coattention}}
    \caption{Encoder blocks. Each yellow area is skipped via residual connection.}
    \label{fig:encoder_blocks}
\end{figure}

EEND-EDA uses a Transformer encoder \cite{vaswani2017attention} without positional encodings (\autoref{fig:transformer}) for $\mathsf{Encoder}$ in \autoref{eq:encoder}.
Given $E_\text{in}\in\mathbb{R}^{D\times T}$, the encoder converts it into $E_\text{out}\in\mathbb{R}^{D\times T}$ as follows: 
\begin{align}
    E'&=\mathsf{LN}\left(E_\text{in}+\mathsf{MA}\left(E_\text{in},E_\text{in},E_\text{in}\right);\vect{\Theta},\vect{\Phi}\right),\\
    E_\text{out}&=\mathsf{LN}\left(E'+\mathsf{FFN}\left(E';\mathsf{\Psi}\right)\right),
\end{align}
where $\vect{\Theta}$, $\vect{\Phi}$, and $\mathsf{\Psi}$ are sets of parameters, and $\mathsf{MA}$ and $\mathsf{FFN}$ denote multi-head scaled dot-product attention and a feed-forward network, respectively, each of which is formulated in the following sections.

\subsubsection{Multi-head scaled dot-product attention}
Given $d_k$-dimensional query $Q\in\mathbb{R}^{d_k\times T}$, key $K\in\mathbb{R}^{d_k\times T}$, and $d_v$-dimensional value $V\in\mathbb{R}^{d_v\times T}$ inputs, multi-head scaled dot-product attention $\mathsf{MA}$ is calculated as
\begin{align}
    \mathsf{MA}\left(Q,K,V;\vect{\Theta},\vect{\Phi}\right)=
    W_O\begin{bmatrix}V^{(1)}A^{(1)\mathsf{T}}\\[-5pt]\vdots\\[-1pt]V^{(h)}A^{(h)\mathsf{T}}\end{bmatrix}+\vect{b}_O\trans{\vect{1}}\in\mathbb{R}^{d_v\times T},\label{eq:multihead_attention}
\end{align}
\vspace{-10pt}
\begin{align}
    A^{(i)}&=\mathsf{softmax}\left(\frac{Q^{(i)\mathsf{T}}K^{(i)}}{\sqrt{d_k/h}}\right)\in\left(0,1\right)^{T\times T},\label{eq:attention}\\
    Q^{(i)}&=W_Q^{(i)}Q+\vect{b}_Q^{(i)}\trans{\vect{1}}\in\mathbb{R}^{\frac{d_k}{h}\times T},\label{eq:q}\\
    K^{(i)}&=W_K^{(i)}K+\vect{b}_K^{(i)}\trans{\vect{1}}\in\mathbb{R}^{\frac{d_k}{h}\times T},\label{eq:k}\\
    V^{(i)}&=W_V^{(i)}V+\vect{b}_V^{(i)}\trans{\vect{1}}\in\mathbb{R}^{\frac{d_v}{h}\times T},\label{eq:v}
\end{align}
where $h$ is the number of heads, $i\in\left\{1,\dots,h\right\}$ is the head index, and $\mathsf{softmax}\left(\cdot\right)$ is the column-wise softmax function.
The set of parameters $\vect{\Theta}$ and $\vect{\Phi}$ are defined as
\begin{align}
    \vect{\Theta}&\coloneqq\bigcup_{1\leq i\leq h}\left\{W_Q^{(i)},\vect{b}_Q^{(i)},W_K^{(i)},\vect{b}_K^{(i)}\right\},\\
    \vect{\Phi}&\coloneqq\left\{W_O,\vect{b}_O\right\}\cup\bigcup_{1\leq i\leq h}\left\{W_V^{(i)},\vect{b}_V^{(i)}\right\}.
\end{align}

\subsubsection{Feed-forward network}
The feed-forward network $\mathsf{FFN}$ consists of two fully connected layers:
\begin{align}
\mathsf{FFN}\left(E';\vect{\Psi}\right)&=\left(W_2\left[W_1E'+\vect{b}_1\trans{\vect{1}}\right]_{+}+\vect{b}_2\trans{\vect{1}}\right),\\
\vect{\Psi}&\coloneqq\left\{W_1,\vect{b}_1,W_2,\vect{b}_2\right\},
\end{align}
where $W_1\in\mathbb{R}^{d_f\times D}$ and $W_2\in\mathbb{R}^{D\times d_f}$ are projection matrices, $\vect{b}_1\in\mathbb{R}^{d_f}$ and $\vect{b}_2\in\mathbb{R}^{D}$ are biases, and $\left[\cdot\right]_{+}$ is the ramp function.

\section{Multi-channel EEND}
To accept multi-channel inputs, we replaced Transformer encoders in EEND-EDA with multi-channel encoders.
In this paper, we investigated two types of encoders: spatio-temporal encoder and co-attention encoder.

\subsection{Spatio-temporal encoder}
The spatio-temporal encoder was originally proposed for speech separation on the basis of distributed microphones \cite{wang2020neural,wang2021continuous}. It uses stacked cross-channel and cross-frame self-attentions in one encoder block, as illustrated in \autoref{fig:spatiotemporal}.
In the encoder, frame-wise $C$-channel embeddings $\vect{E}_\text{in}=\left(\vect{e}_{\text{in},t,c}\right)_{t,c}\in\mathbb{R}^{D\times T\times C}$, where $\vect{e}_{\text{in},t,c}\in\mathbb{R}^{D}$, are first converted to the same shape of tensor $\vect{E}'=\left(\vect{e}'_{t,c}\right)_{t,c}\in\mathbb{R}^{D\times T\times C}$ using cross-channel self-attention as
\begin{align}
    \left[\vect{e}'_{t,1},\dots,\vect{e}'_{t,C}\right]&=\mathsf{LN}\left(E_{\text{in},t}+\mathsf{MA}\left(E_{\text{in},t},E_{\text{in},t},E_{\text{in},t};\vect{\Theta},\vect{\Phi}\right)\right),\label{eq:cross_channel_msa}\\
    E_{\text{in},t}&\coloneqq\left[\vect{e}_{\text{in},t,1},\dots,\vect{e}_{\text{in},t,C}\right].\label{eq:cross_frame_msa}
\end{align}
The tensor $\vect{E}'$ is then converted to $\vect{E}_\text{out}=\left(\vect{e}_{\text{out},t,c}\right)_{t,c}\in\mathbb{R}^{D\times T\times C}$ by cross-frame self-attention as
\begin{align}
    \left[\vect{e}_{\text{out},1,c},\dots,\vect{e}_{\text{out},T,c}\right]&=\mathsf{LN}\left(E'_c+\mathsf{MA}\left(E'_c,E'_c,E'_c;\vect{\Theta}',\vect{\Phi}'\right)\right),\label{eq:cross_frame}\\
    E'_c&\coloneqq\left[\vect{e}'_{1,c},\dots,\vect{e}'_{T,c}\right].\label{eq:def_ec}
\end{align}
In the final encoder block, cross-frame self-attention is calculated over the embeddings that are averaged across channels to form $E^{(N)}$ in \autoref{eq:attractors}, \ie, the following are used instead of \autoref{eq:cross_frame} and \autoref{eq:def_ec} as
\begin{align}
   E^{(N)}&=\mathsf{LN}\left(E'+\mathsf{MA}\left(E',E',E';\vect{\Theta}',\vect{\Phi}'\right)\right),\\
    E'&\coloneqq\frac{1}{C}\sum_{c=1}^{C}E'_c\label{eq:average_ec}
\end{align}
to calculate speech activities using \autoref{eq:attractors} and \autoref{eq:posteriors}.
All calculations using \autoref{eq:cross_channel_msa}--\autoref{eq:average_ec} do not involve a specific number of channels or microphone geometry, which makes this encoder independent of the number and geometry of microphones.
Note that we did not include feed-forward networks $\mathsf{FFN}$ in this encoder following previous studies \cite{wang2020neural,wang2021continuous} because we observed performance degradation.

\subsection{Co-attention encoder}
The spatio-temporal encoder includes cross-channel self-attention, the performance of which highly depends on the number of channels.
Therefore, we also propose an encoder based only on cross-frame attention, which is characterized by the use of co-attention.
The encoder accepts two inputs: frame-wise embeddings $E_\text{in}\in\mathbb{R}^{D\times T}$ and frame-channel-wise embeddings $\vect{P}_\text{in}=[P_{\text{in},1},\dots,P_{\text{in},C}\mid P_{\text{in},c}\in\mathbb{R}^{D'\times T}]$.
The proposed encoder converts these inputs to $E_\text{out}\in\mathbb{R}^{D\times T}$ and $\vect{P}_\text{out}=[P_{\text{out},1},\dots,P_{\text{out},C}\mid P_{\text{out},c}\in\mathbb{R}^{D'\times T}]$ as follows:
\begin{align}
    E'&=\mathsf{LN}\left(E_\text{in}+\mathsf{MCA}\left(\vect{P}_\text{in},\vect{P}_\text{in},E_\text{in};\vect{\Theta}_P,\vect{\Phi}_E\right)\right),\label{eq:mono_mca}\\
    E''&=\mathsf{LN}\left(E'+\mathsf{MA}\left(E',E',E';\vect{\Theta}'_E,\vect{\Phi}'_E\right)\right)\label{eq:mono_msa}\\
    E_\text{out}&=\mathsf{LN}\left(E''+\mathsf{FFN}\left(E'';\vect{\Psi}_E\right)\right),\label{eq:mono_ffn}\\
    P'_c&=\mathsf{LN}\left(P_{\text{in},c}+\mathsf{MCA}\left(\vect{P}_\text{in},\vect{P}_\text{in},P_{\text{in},c};\vect{\Theta}_P,\vect{\Phi}_P\right)\right),\label{eq:multi_mca}\\
    P_{\text{out},\text{c}}&=\mathsf{LN}\left(P'_c+\mathsf{FFN}\left(P'_c;\vect{\Psi}_P\right)\right),\label{eq:multi_ffn}
\end{align}
where $\vect{\Theta}_P$, $\vect{\Phi}_E$, $\vect{\Theta}'_E$, $\vect{\Phi}'_E$, $\vect{\Psi}_E$, $\vect{\Psi}_P$, $\vect{\Phi}_P$, and $\vect{\Psi}_P$ are the sets of parameters.
The single-channel input $E_\text{in}$ is converted by multi-head co-attention $\mathsf{MCA}$ in \autoref{eq:mono_mca}, multi-head attention $\mathsf{MA}$ in \autoref{eq:mono_msa}, and feed-forward network $\mathsf{FFN}$ in \autoref{eq:mono_ffn}.
Each channel in the multi-channel input $\vect{P}_\text{in}$ is first converted by $\mathsf{MCA}$ in \autoref{eq:multi_mca}, the attention weights of which are shared with those in \autoref{eq:mono_mca}, then processed using $\mathsf{FFN}$ in \autoref{eq:multi_ffn}.

Multi-head co-attention $\mathsf{MCA}$ is similar to $\mathsf{MA}$ in \autoref{eq:multihead_attention}, but the attention weights are calculated using multi-channel inputs as
\begin{gather}
    \mathsf{MCA}\left(\vect{Q},\vect{K},V;\vect{\Theta},\vect{\Phi}\right)=
    W_O\begin{bmatrix}V^{(1)}A^{(1)\mathsf{T}}\\[-5pt]\vdots\\[-1pt]V^{(h)}A^{(h)\mathsf{T}}\end{bmatrix}+\vect{b}_O\trans{\vect{1}}\in\mathbb{R}^{d_v\times T},\\
    A^{(i)}=\mathsf{softmax}\left(\frac{\left[Q^{(i)\mathsf{T}}_1,\dots,Q^{(i)\mathsf{T}}_C\right]\trans{\left[K^{(i)\mathsf{T}}_1,\dots,K^{(i)\mathsf{T}}_C\right]}}{\sqrt{CD/h}}\right).
\end{gather}
Here, $Q_c^{(i)}$ and $K_c^{(i)}$ for $c\in\left\{1,\dots,C\right\}$ are calculated using \autoref{eq:q} and \autoref{eq:k} for each channel, and $V^{(i)}$ are calculated using \autoref{eq:v}.
Note that the parameter sets $\vect{\Theta}$ and $\vect{\Phi}$ are shared among channels.

After the final encoder block, two outputs are concatenated as
\begin{align}
    E^{(N)}=\begin{bmatrix}E_{\text{out}}\\\frac{1}{C}\sum_{c=1}^{C}P_{\text{out},c}\end{bmatrix}\in\mathbb{R}^{(D+D')\times T}
\end{align}
to calculate speech activities using \autoref{eq:attractors} and \autoref{eq:posteriors}.

\subsection{Domain Adaptation}
EEND performance can be improved by domain adaptation using real recordings.
However, the number of real recordings is usually limited, and even more the case when distributed microphones are used.
Therefore, it would be useful if multi-channel EEND can be adapted to the target domain only with single-channel recordings.
To ensure that adaptation using only single-channel recordings does not lose the ability to benefit from multi-channel recordings, we propose to update only the channel-invariant part of the model.
For the spatio-temporal encoder, we freeze the parameters of cross-channel self-attention $\vect{\Theta}$ and $\vect{\Phi}$ in \autoref{eq:cross_channel_msa}.
For the co-attention encoder, we freeze the parameters related to multi-channel processing: $\vect{\Theta}_P$ in \autoref{eq:mono_mca} and \autoref{eq:multi_mca}, $\vect{\Phi}_P$ in \autoref{eq:multi_mca}, and $\vect{\Psi}_P$ in \autoref{eq:multi_ffn}.

\begin{table}[t]
    \centering
    \caption{Two-speaker conversational datasets.}
    \vspace{-6pt}
    \setlength{\tabcolsep}{3pt}
    \label{tbl:dataset}
    \resizebox{\linewidth}{!}{%
    \begin{tabular}{@{}lccccrr@{}}
        \toprule
        &Conver-&&&&\multicolumn{1}{c}{Average}&\multicolumn{1}{c@{}}{Overlap}\\
        Dataset & sation&Record&\#Mic & \#Session&\multicolumn{1}{c}{duration}&\multicolumn{1}{c}{ratio}\\\midrule
        SRE+SWBD-train & Simulated&Simulated&10&20,000&\SI{88.7}{\second}&\SI{34.1}{\percent} \\
        SRE+SWBD-eval & Simulated&Simulated&10&500&\SI{88.1}{\second}&\SI{34.6}{\percent}\\
        SRE+SWBD-eval-hybrid & Simulated&Simulated&10&500&\SI{88.1}{\second}&\SI{34.6}{\percent}\\
        CSJ-train & Simulated &Recorded&9&100&\SI{113.5}{\second}&\SI{11.0}{\percent}\\
        CSJ-eval & Simulated&Recorded&9&100&\SI{102.2}{\second}&\SI{9.6}{\percent}\\
        CSJ-dialog & Real&Recorded&9&58&\SI{755.2}{\second}&\SI{17.3}{\percent}\\
        \bottomrule
    \end{tabular}%
    }
\end{table}

\begin{table}[t]
    \centering
    \caption{DERs on SRE+SWBD-eval and SRE+SWBD-hybrid.}
    \vspace{-6pt}
    \label{tbl:results_SRE+SWBD_eval}
    \setlength{\tabcolsep}{3pt}
    \begin{threeparttable}
    \resizebox{\linewidth}{!}{%
    \begin{tabular}{@{}lcccccccccc@{}}
        \toprule
        &\multicolumn{5}{c}{SRE+SWBD-eval}&\multicolumn{5}{c}{SRE+SWBD-eval-hybrid}\\\cmidrule(l{3pt}r{3pt}){2-6}\cmidrule(l{3pt}){7-11}
        Method&1ch&2ch&4ch&6ch&10ch&1ch&2ch&4ch&6ch&10ch\\\midrule
        1ch + posterior avg.& 5.13& 4.60& 4.31&4.19&4.10& 6.07& 5.68&5.42&5.38&\textbf{5.33}\\\midrule
        Spatio-temporal&32.86&2.97&\textbf{1.49}&\textbf{1.19}&\textbf{1.03}&34.73&10.60&8.65&8.36&8.21\\
        Spatio-temporal\tnote{\dag}& 6.34&3.02&1.56&1.28&1.07& 8.11&8.23&6.98&6.72&6.40\\\midrule
        Co-attention&7.23&2.83&1.85&1.59&1.50&9.03&7.53&6.82&6.51&6.65\\
        Co-attention\tnote{\dag}&\textbf{4.68}&\textbf{2.52}&1.71&1.40&1.23&\textbf{5.73}&\textbf{5.34}&\textbf{5.05}&\textbf{5.18}&5.35\\
        \bottomrule
    \end{tabular}%
    }
    \begin{tablenotes}
        \footnotesize
        \item[\dag] Channel dropout was used during training.
    \end{tablenotes}
    \end{threeparttable}
\end{table}

\begin{figure*}[t!]
\begin{minipage}{0.61\linewidth}
    \centering
    \captionsetup{type=table}
    \captionof{table}{DERs on CSJ-eval and CSJ-dialog.}
    \vspace{-6pt}
    \label{tbl:results_csj}
    \begin{threeparttable}
    \resizebox{\linewidth}{!}{%
    \begin{tabular}{@{}lccccccccccc@{}}
        \toprule
        &&\multicolumn{5}{c}{CSJ-eval}&\multicolumn{5}{c}{CSJ-dialog}\\\cmidrule(lr){3-7}\cmidrule(l){8-12}
        Method&Adapt&1ch&2ch&4ch&6ch&9ch&1ch&2ch&4ch&6ch&9ch\\\midrule
        1ch + posterior avg. &None& 11.17&9.44&8.94&8.89&8.44&28.15&26.01&25.56&24.74&24.87\\
        1ch + posterior avg.&1ch&3.27&2.31&2.25&2.05&1.75&22.56&20.82&20.34&19.68&20.25\\\midrule
        Spatio-temporal &None& 10.98&10.20&4.29&3.27&2.63&36.13&45.19&36.48&37.14&37.63\\
        Spatio-temporal &1ch& 3.44&1.60&1.34&1.07&1.13&\textbf{20.06}&20.02&17.83&16.19&19.74\\
        Spatio-temporal&1ch\tnote{\ddag}&3.64&1.78&1.64&1.27&1.32&20.57&19.02&17.37&15.49&18.70\\
        Spatio-temporal &4ch& 3.82&\textbf{1.06}&0.61&0.43&\textbf{0.39}&21.01&\textbf{15.87}&\textbf{14.21}&15.71&14.20\\\midrule
        Co-attention&None& 9.49&3.36&1.42&1.40&0.94&27.96&22.52&19.37&18.23&17.99\\
        Co-attention&1ch& \textbf{2.75}&1.41&0.75&0.63&0.52&23.49&22.83&20.70&17.59&15.77\\
        Co-attention&1ch\tnote{\ddag}& 3.26&1.46&0.68&0.48&0.42&22.45&17.90&15.53&14.34&14.05\\
        Co-attention&4ch& 3.31&1.19&\textbf{0.57}&\textbf{0.40}&\textbf{0.39}&21.42&17.51&14.95&\textbf{14.21}&\textbf{13.87}\\
        \bottomrule
    \end{tabular}%
    }
    \begin{tablenotes}
        \footnotesize
        \item[\ddag] Adapted only channel-invariant part of each model.
    \end{tablenotes}
    \end{threeparttable}
\end{minipage}%
\hfill%
\begin{minipage}{0.27\linewidth}
    \centering
    \includegraphics[width=\linewidth]{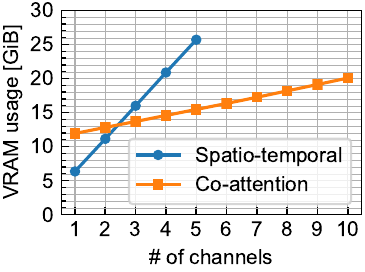}
    \caption{VRAM usage during training with $T=500$ and batch size of 64.}
    \label{fig:vram_usage}
\end{minipage}%
\end{figure*}

\section{Experiment}\label{sec:experiment}
\subsection{Datasets}\label{sec:dataset}
For the experiments, we created three fully simulated two-speaker conversational datasets using NIST Speaker Recognition Evaluation (2004--2006, 2008) (SRE), Switchboard-2 (Phase I--I\nobreak\hspace{-.1em}I\nobreak\hspace{-.1em}I), and Switchboard Cellular (Part 1, 2) (SWBD).
To emulate a reverberant environment, we generated room impulse responses (RIRs) using gpuRIR \cite{diaz2021gpurir}.
Following the procedure in \cite{ko2017study}, we sampled 200 rooms for each of the three room sizes: small, medium, and large.
In each room, a table was randomly placed, 10 speakers were randomly placed around the table, and 10 microphones were randomly placed on the table.
To create SRE+SWBD-train and SRE+SWBD-eval, two-speaker conversations were simulated following \cite{fujita2019end1} then RIRs of the randomly selected room and two speaker positions were convolved to obtain a 10-channel mixture.
MUSAN corpus \cite{snyder2015musan} was also used to add noise to each mixture.
SRE+SWBD-eval-hybrid was created using the same utterances in SRE+SWBD-eval, but two speakers were placed at the same position.
This dataset was designed to mimic the part of hybrid meetings, in which multiple speakers' utterances are played from a single loudspeaker.

We also prepared three real-recorded datasets on the basis of the corpus of spontaneous Japanese (CSJ) \cite{maekawa2003corpus}: CSJ-train, CSJ-eval, and CSJ-dialog.
For CSJ-train and CSJ-eval, 100 two-speaker conversations were simulated using single-speaker recordings in the CSJ training and evaluation sets, respectively.
For CSJ-dialog, we directly used the dialog portion of CSJ.
To record each session, we distributed nine smartphone devices on a tabletop in a meeting room and four loudspeakers around the table.
We played back two speakers' utterances from two of the four loudspeakers that were randomly selected and recorded them on the smartphone devices.
Recorded signals were roughly synchronized to maximize the correlation coefficient and neither clock drift nor frame dropping was compensated.

All the experiments were based on two-speaker mixtures because our scope was investigating multi-channel diarization.
Note that EEND-EDA can also be used when the number of speakers is unknown \cite{horiguchi2020endtoend,horiguchi2021encoder}.

\subsection{Settings}
\label{sec:setting}
As inputs to a single-channel baseline model \cite{horiguchi2020endtoend,horiguchi2021encoder}, 23-dimensional log-mel filterbanks were extracted for each \SI{10}{\ms} followed by splicing ($\pm 7$ frames) and subsampling by factor of 10, resulting in 345-dimensional features for each \SI{100}{\ms}.
For the spatio-temporal model, we extracted features from each channel in the same manner.
For the co-attention model, the 345-dimensional features were averaged across channels to be used as the single-channel input.
As the multi-channel input, the log-mel filterbanks of $\pm 7$ frames were averaged followed by subsampling; thus, a 23-dimensional feature was obtained for each \SI{100}{\ms}.
We set the embedding dimensionalities as $D=256$ and $D'=64$, \ie, 345-dimensional features were first converted to 256 dimensional via \autoref{eq:change_dimension} and 23-dimensional features were converted to 64 dimensional in the same manner.
For each model, the four encoder blocks illustrated in \autoref{fig:encoder_blocks} were stacked.

Each model was first trained on SRE+SWBD-train for 500 epochs with the Adam optimizer \cite{kingma2015adam} using Noam scheduler \cite{vaswani2017attention} with 100,000 warm-up steps.
At each iteration, four of ten channels were randomly selected and used for training.
The models were then evaluated on SRE+SWBD-eval and SRE+SWBD-eval-hybrid using $\{1,2,4,6,10\}$-channel inputs.
Each model was further adapted to CSJ-train for 100 epochs with the Adam optimizer with a fixed learning rate of $1\times10^{-5}$.
The adapted models were evaluated on CSJ-eval and CSJ-dialog using $\{1,2,4,6,9\}$-channel inputs.
To evaluate the conventional EEND-EDA \cite{horiguchi2020endtoend,horiguchi2021encoder} with multi-channel inputs, we first found the optimal speaker permutation between results from each channel by using correlation coefficients of posteriors and then averaged the posteriors among channels.
To prevent the models from being overly dependent on spatial information, we also introduce channel dropout, in which multi-channel inputs are randomly dropped to be a single channel.
The ratio of channel dropout was set to 0.1.
Each method was evaluated using diarization error rates (DERs) with \SI{0.25}{\second} of collar tolerance.

\subsection{Results}
\autoref{tbl:results_SRE+SWBD_eval} shows the DERs on SRE+SWBD-eval and SRE+SWBD-eval-hybrid.
From the results on SRE+SWBD-eval, both spatio-temporal and co-attention models outperformed the single-channel model with posterior averaging.
Comparing the two multi-channel models, the spatio-temporal model significantly degraded DER with single-channel inputs.
Channel dropout eased the situation, but the co-attention model still outperformed the spatio-temporal model when the number of channels was small.
In the evaluation of SRE+SWBD-eval-hybrid, the co-attention model always achieved the same or better DERs than the single-channel model.
This means that the lack of spatial information does not lead to degradation in diarization performance in the co-attention model because it does not rely on cross-channel self-attention.

\autoref{tbl:results_csj} shows the DERs on CSJ-eval and CSJ-dialog.
The evaluation was based on the models trained using channel dropout.
Without adaptation, we can see that the co-attention model generalized well.
The performance of all models improved through adaptation, regardless of whether the data used for adaptation were 1ch or 4ch.
Of course, both spatio-temporal and co-attention models can benefit more from 4ch adaptation; however, it is worth mentioning that they can still utilize spatial information provided by multi-channel inputs even if only 1ch recordings are used for adaptation.
By freezing the parameters related to the calculation across channels during 1ch adaptation, the DERs of the co-attention model were reduced especially when four or more microphones were used, while those of the spatio-temporal model were not so improved.

Finally, we show the peak VRAM usage with $T=500$ and batch size of 64 in \autoref{fig:vram_usage}.
VRAM usage of the co-attention model increased more slowly than the spatio-temporal model as the number of microphones increased because the multi-channel processing part is based on layers with a lower number of units.
Thus, the co-attention model can be trained using a larger number of channels.

\section{Conclusion}\label{sec:conclusion}
In this paper, we proposed a multi-channel end-to-end neural diarization method based on distributed microphones.
We replaced Transformer encoders in the conventional EEND with two types of multi-channel encoders.
Each showed better DERs with multi-channel inputs than the conventional EEND on both simulated and real-recorded datasets.
We also proposed a model adaptation method using only single-channel recordings, and achieved comparable DERs as when using multi-channel recordings.

\bibliographystyle{IEEEbib-abbrev}
\bibliography{refs}

\end{document}